\documentclass[final,1p,times,authoryear]{elsarticle}
\usepackage{amssymb}
\usepackage{amsmath}
\usepackage{cleveref}
\crefname{figure}{Fig.}{Figs.}
\begin{document}
\begin{frontmatter}
\title{Unsupervised Denoising of Diffusion-Weighted Images with Bias and Variance Corrected Noise Modeling} %% Article title

\author[label1,label2]{Jine Xie\fnref{fn1}} %% Author name
\author[label1,label3]{Zhicheng Zhang\fnref{fn1}}
\author[label1,label4]{Yunwei Chen}
\author[label1,label2]{Yanqiu Feng}
\author[label1,label2]{Xinyuan Zhang\corref{cor1}}

\affiliation[label1]{organization={School of Biomedical Engineering, Southern Medical University},
            %addressline={},
            city={Guangzhou},
            %postcode={},
            %state={},
            country={China}}
            
\affiliation[label2]{organization={Guangdong Provincial Key Laboratory of Medical Image Processing, Southern Medical University},
            %addressline={},
            city={Guangzhou},
            %postcode={510515},
            %state={},
            country={China}}
            
\affiliation[label3]{organization={JancsiLab, JancsiTech},
            %addressline={},
            city={Hong Kong},
            %postcode={},
            %state={},
            country={China}}

\affiliation[label4]{organization={Department of Radiation Oncology, Sun Yat-sen University Cancer Center},
            %addressline={},
            city={Guangzhou},
            %postcode={},
            %state={},
            country={China}}
            
\cortext[cor1]{Corresponding author. e-mail:zhangxyn@smu.edu.cn}        
\fntext[fn1]{These authors contributed equally to this work.}

%% Abstract
\begin{abstract}
Diffusion magnetic resonance imaging (dMRI) plays a vital role in both clinical diagnostics and neuroscience research. However, its inherently low signal-to-noise ratio (SNR), especially under high diffusion weighting, significantly degrades image quality and impairs downstream analysis. Recent self-supervised and unsupervised denoising methods offer a practical solution by enhancing image quality without requiring clean references. However, most of these methods do not explicitly account for the non-Gaussian noise characteristics commonly present in dMRI magnitude data during the supervised learning process, potentially leading to systematic bias and heteroscedastic variance, particularly under low-SNR conditions. To overcome this limitation, we introduce noise-corrected training objectives that explicitly model Rician statistics. Specifically, we propose two alternative loss functions: one derived from the first-order moment to remove mean bias, and another from the second-order moment to correct squared-signal bias. Both losses include adaptive weighting to account for variance heterogeneity and can be used without changing the network architecture. These objectives are instantiated in an image-specific, unsupervised Deep Image Prior (DIP) framework. Comprehensive experiments on simulated and in-vivo dMRI show that the proposed losses effectively reduce Rician bias and suppress noise fluctuations, yielding higher image quality and more reliable diffusion metrics than state-of-the-art denoising baselines. These results underscore the importance of bias- and variance-aware noise modeling for robust dMRI analysis under low-SNR conditions.
\end{abstract}

%%Graphical abstract 图文摘要


%% Keywords
\begin{keyword}
Diffusion magnetic resonance imaging \sep noise-corrected model \sep image denoising \sep Rician noise \sep deep image prior.
\end{keyword}

\end{frontmatter}


\section{Introduction}
\label{sec1}
%% Labels are used to cross-reference an item using\ref command.
Diffusion magnetic resonance imaging (dMRI) is a non-invasive technique that probes the diffusion of water molecules to reveal tissue microstructure \citep{LeBihan2003FunctionalArchitecture}. To quantify these microstructural features, researchers have developed a spectrum of models, from phenomenological methods like diffusion tensor imaging (DTI) \citep{Mori2006DTIPrinciples} and diffusion kurtosis imaging (DKI) \citep{Jensen2005DiffusionalKurtosis} to biophysical approaches such as neurite orientation dispersion and density imaging (NODDI) \citep{Zhang2012NODDI} as well as soma and neurite density imaging (SANDI) \citep{Palombo2020SANDI}. These techniques are now integral to clinical and neuroscience studies, facilitating disease diagnosis and enhancing our understanding of brain development and pathology.

Despite these advances, the reliability of dMRI analysis remains fundamentally constrained by the low signal-to-noise ratio (SNR) of diffusion-weighted (DW) images, which will be exacerbated at high b-values or high spatial resolutions. Low SNR not only impairs visual image quality but also undermines the accuracy and robustness of subsequent diffusion parameter estimation. While strategies such as stronger magnetic fields, optimized RF coils, or repeated acquisitions can improve SNR, they are costly and often introduce motion artifacts due to prolonged scan times. As a more practical alternative, post-processing denoising algorithms have been extensively explored \citep{Manzano2024DenoisingConsiderations}.

Traditional denoising methods for dMRI leverage redundancies in spatial (x-space) or diffusion (q-space) dimensions to suppress noise. Common strategies include transforming image patches into an eigen-domain for spectral filtering \citep{Veraart2016MPPCA,Ma2020DenoiseVST,Zhang2017HOSVD,Moeller2021NORDIC,Ye2025TwoStepPCA, Huynh2024OptimalShrinkage} or using neighborhood matching in the image domain \citep{Chen2016XQNLM,Chen2019NoiseReduction,WiestDaessle2007NLMVariants}. Among the above methods, some are designed to process magnitude data, while others require access to complex-valued data. Magnitude-domain denoising typically handles Rician-distributed noise via variance-stabilizing transforms (VST) applied before and after denoising \citep{Foi2011NoiseEstimation}, or by correcting residual bias with moment-based approaches on the denoised data \citep{Koay2006AnalyticallyExact}. In contrast, complex-domain denoising operates directly on the real and imaginary channels, where the noise follows a Gaussian distribution, eliminating the need for Rician-specific corrections. While complex-domain methods generally outperform magnitude-based approaches, their clinical application is limited because routine reconstructions typically provide only magnitude data.

More recently, deep learning methods, particularly convolutional neural networks (CNNs) such as DnCNN \citep{Zhang2017BeyondGaussian}, have demonstrated superior denoising performance. Supervised learning frameworks \citep{Tian2020DeepDTI,Dou2025MRI,Jurek2023SupervisedDenoising}, however, require clean ground-truth references, which are infeasible to obtain in practice since multiple repeated acquisitions and complex-domain averaging would be necessary to reduce both noise variance and Rician bias \citep{Moeller2021NORDIC}. To circumvent this limitation, self-supervised and unsupervised approaches have emerged, leveraging spatial and q-space redundancies without reference data. Initially, Patch2Self exploited the statistical independence of noise and J-invariance to predict the DW signal in one direction from patches in other directions, enabling self-supervised learning\citep{Fadnavis2020Patch2Self}. Patch2Self2 further accelerated the training process of Patch2Self by using matrix sketching techniques \citep{Fadnavis2024Patch2Self2}. Other deep learning-based J-invariant strategies, such as Replace2Self \citep{Wu2025Replace2Self}, SAN2N \citep{Jiang2024SelfSupervisedNoise2Noise}, and MD-S2S \citep{Kang2024SelfSupervisedMultidimensional}, achieve self-supervision by constructing noisy image pairs using masking or sampling across x- and q-space. Meanwhile, the generative approach DDM2 performs self-supervised dMRI denoising through noise estimation to guide forward-state matching, followed by reconstruction via reverse diffusion \citep{Xiang2023DDM2}. Tian et al. also introduced a self-supervised framework incorporating dMRI models as constraints, either by using DTI to generate paired low-SNR and high-SNR images \citep{Tian2022SDnDTI} or by producing synthetic data from network-parameterized models and enforcing consistency with the acquired data \citep{Li2024DIMOND}. One study also explores an unsupervised scheme based on the Deep Image Prior (DIP) paradigm \citep{Lin2020DeepImagePriorDWI}. While these self-supervised or unsupervised approaches have achieved notable success, they typically focus on constructing data pairs by leveraging multidimensional data redundancies, without explicitly accounting for the underlying non-Gaussian noise characteristics. Consequently, this oversight may lead to residual bias, particularly under low-SNR conditions. 

Although complex-domain deep learning \citep{Dou2025MRI} can in principle overcome this issue by exploiting Gaussian-distributed noise, they share the same limitation as traditional methods: complex-valued data are rarely available in routine clinical practice, where only magnitude reconstructions are typically provided. Previous studies have shown that first- and second-order moment–based fitting can reduce MRI parameter estimation bias under Rician/non-central chi noise \citep{Feng2013ImprovedRelaxometry, Guo2022JointFramework}. Motivated by these insights and addressing the limitations of current self- and unsupervised loss formulations, in this work, we propose two novel noise-corrected loss functions for magnitude dMRI denoising, which explicitly model Rician statistics. Specifically, our formulation applies first-order moment correction to eliminate mean bias and second-order moment correction to address squared-signal bias. In addition, adaptive weighting is introduced to account for voxel-wise variance heterogeneity, resulting in a variance-normalized objective well aligned with the signal-dependent noise characteristics of magnitude dMRI. We evaluate these two objectives within the DIP framework \citep{Lempitsky2018DeepImagePrior}, whose simplicity, unsupervised nature, and minimal reliance on training data make it a transparent and effective proof-of-concept for validating new loss functions. By removing confounding factors such as network complexity and dataset size, DIP highlights the specific benefits of bias- and variance-corrected objectives in magnitude dMRI denoising. Additionally, experiments on simulated and in-vivo dMRI data demonstrate that our method outperforms state-of-the-art denoising techniques, particularly under severe noise conditions. In summary, the main contributions of this work are three-fold:
\begin{itemize}
    \item Two more targeted loss functions that explicitly model Rician noise by jointly correcting bias and accounting for variance heterogeneity are proposed.
    \item An unsupervised validation framework based on the DIP framework is established, enabling transparent assessment of noise-corrected objectives without reliance on large training datasets.
    \item Through experiments on simulated and in-vivo dMRI data, we demonstrate that our approach delivers superior denoising performance and more accurate diffusion parameter estimation compared with state-of-the-art methods.
\end{itemize}

\section{Theory}
\label{sec2}
\subsection{Definition of Notations}
\label{sec2subsec1}
In this paper, lowercase letters (such as $x$) represent scalar quantities, while bold lowercase letters (such as $\boldsymbol{x}$) denote vectors. Uppercase letters (such as $X$) represent two-dimensional matrices, and bold uppercase letters (such as $\boldsymbol{X}$) are used to denote tensors, which correspond to three-dimensional or higher-dimensional structures. $\boldsymbol{X}_i$ denotes an individual element or voxel from a tensor $\boldsymbol{X}$ (such as those in dMRI data).

\subsection{General Loss Formulation for Image Denoising}
\label{sec2subsec2}
Let the observed noisy image be expressed as:
\begin{equation}
Y = X + n,\quad n \sim \mathcal{N}(0, \sigma)
\end{equation}
where $X$ is the unknown clean image, $Y$ is the acquired noisy image, and $n$ is Gaussian noise with zero mean and standard deviation (SD) $\sigma$.

Generally, image denoising is formulated under the Gaussian noise assumption as an optimization problem that minimizes a data fidelity term combined with a regularization term:
\begin{equation}
\hat{X} = \min_X \left\{ \|Y - X\|_2^2 + \lambda \mathcal{R}(X) \right\}
\end{equation}
where $\hat{X}$ denotes the estimated clean image, $\mathcal{R}(\cdot)$ is a regularization term that enforces prior knowledge such as smoothness or sparsity. Typical choices for $\mathcal{R}(\cdot)$ include low-rank matrix or tensor priors, total variation (TV), or data-driven priors, depending on the desired structural properties of the restored image.

\subsection{Loss Formulation in DIP}
\label{sec2subsec3}
In the Deep Image Prior (DIP) framework, denoising is performed by optimizing a convolutional network $f_{\theta}(\cdot)$ to fit the noisy image $Y$. Unlike classical hand-crafted prior methods or other data-driven deep models, DIP exploits the network architecture itself as an implicit prior for image restoration.

The optimization can be written as:
\begin{equation}
\hat{\boldsymbol{\theta}} = \arg\min_{\boldsymbol{\theta}} \|f_{\boldsymbol{\theta}}(Z) - Y\|_2^2, \quad
\hat{X} = f_{\hat{\boldsymbol{\theta}}}(Z)
\end{equation}
where $Z$ is a fixed random noise map, $f_{\theta}(\cdot)$ is a neural network with learnable parameters $\boldsymbol{\theta}$, and $\hat{X}$ is the denoised output. The optimization process uses the noisy image ($Y$) as the regression target, with random noise map ($Z$) as the network input, and applies reparameterization $X=f_{\boldsymbol{\theta}}(Z)$ as a constraint, guiding the network to estimate a noise-free image.

\subsection{Bias and Variance Analysis under Rician Noise}
\label{sec2subsec4}
Unlike Gaussian noise, the dMRI magnitude signal $y$ follows a Rician distribution after taking the square root of the complex data. Its Probability Density Function (PDF), which depends on the original noise-free signal $x$, is given as \citep{Gudbjartsson1995Rician}:
\begin{equation}
p(y \mid x; \sigma) = \frac{y}{\sigma^2} \exp\left(-\frac{x^2 + y^2}{2\sigma^2}\right) I_0\left(\frac{xy}{\sigma^2}\right)
\end{equation}
where $I_0 (\cdot)$ is the modified Bessel function of the first kind (zero order) and $\sigma$ denotes the SD of the Gaussian noise in the real and imaginary images (assumed equal).

\subsubsection{Expectation and Variance of $y$}
\label{sec2sub4subsubsec1}
The expectation of $y$ (i.e., first-moment) is \citep{Koay2006AnalyticallyExact,Raya2010T2Measurement}:
\begin{equation}
E(y \mid x; \sigma) = \frac{1}{2\sigma^2} e^{-\frac{x^2}{4\sigma^2}} \sqrt{\frac{\pi}{2}} \, \sigma \,
\left[ (x^2 + 2\sigma^2) I_0\!\left(\frac{x^2}{4\sigma^2}\right) + x^2 I_1\!\left(\frac{x^2}{4\sigma^2}\right) \right]
\label{Eq:expectation_of_y}
\end{equation}
where $I_1 (\cdot)$ is the modified Bessel function of the first kind (first order). Unlike Gaussian distribution, the expectation of $y$ deviates from the underlying noise-free signal $x$ and depends on both the value of $x$ and $\sigma$. This deviation becomes more pronounced as the SNR decreases.

The variance of $y$ is expressed as \citep{Koay2006AnalyticallyExact,Coupe2010RobustRician}:
\begin{equation}
\operatorname{Var}(y \mid x; \sigma) = \sigma^2 \Biggl( 2 + \frac{x^2}{\sigma^2} - \frac{\pi}{8} e^{-\frac{x^2}{2\sigma^2}}
\Biggl[ \biggl( 2 + \frac{x^2}{\sigma^2} \biggr) I_0\!\biggl(\frac{x^2}{4\sigma^2}\biggr) +
\frac{x^2}{\sigma^2} I_1\!\biggl(\frac{x^2}{4\sigma^2}\biggr) \Biggr]^2 \Biggr)
\label{Eq:variance_of_y}
\end{equation}

The variance is also signal-dependent, typically ranging from $0.43\sigma^2$ to $\sigma^2$.

\subsubsection{Implication}
\label{sec2sub4subsubsec2}
For denoising images affected by Rician noise, the estimator $\min\limits_{\boldsymbol{\theta}} \| f_{\boldsymbol{\theta}}(Z) - Y \|_2^2$ in DIP is biased, as the network output $f_{\boldsymbol{\theta}}(Z)$ corresponds to the expectation of $Y$, i.e., $E(f_{\boldsymbol{\theta}}(Z))=E(Y)$. It can reduce noise-induced fluctuations but cannot eliminate the noise bias. Moreover, the variance of $Y$ is heteroscedastic, i.e., it depends on the local signal intensity. This implies that different spatial locations, associated with different signal values, exhibit different variances. As a result, the contribution of each location to the optimization is uneven, and the solution obtained by minimizing $\|f_{\boldsymbol{\theta}}(Z) - Y \|_2^2$ is not globally optimal.

In theory, an estimator should satisfy two key conditions: (i) the expected fitting error is zero when the estimate matches the true signal, and (ii) the variance of the fitting error remains constant (i.e., homoscedastic) across all signal levels. However, Rician noise violates these conditions, causing the resulting estimator to exhibit signal-dependent bias and variance. This leads to suboptimal reconstructions unless these factors are explicitly addressed in the modeling or optimization process.

\subsubsection{Expectation and Variance of $y^2$}
\label{sec2sub4subsubsec3}
Similar to the analysis of $y$, the expectation of $y^2$ (i.e., second-moment ) \citep{Koay2006AnalyticallyExact,Gudbjartsson1995Rician} can be expressed simply as:
\begin{equation}
E(y^2|x;\sigma) = x^2+2{\sigma}^2
\label{Eq:expectation_of_y^2}
\end{equation}

Meanwhile, we derive the variance of $y^2$ as:
\begin{equation}
Var(y^2|x;\sigma) = 4{\sigma}^2(x^2+{\sigma}^2)
\label{Eq:variance_of_y^2}
\end{equation}

The bias of $E(y^2 |x;\sigma)$ from $x^2$ equals the constant term $2{\sigma}^2$, while the variance of $y^2$ depends on both the underlying signal and the noise level. Compared with $y$, the heteroscedasticity of $y^2$ is more pronounced, making the impact of noise more severe.

\section{Materials and methods}
\label{sec3}
\subsection{ Rician Noise-Corrected Loss Function}
\label{sec3subsec1}
Building upon the above analysis, we propose two noise-correction loss functions that explicitly address both bias and variance in the presence of Rician noise. Specifically, we design models based on the first- and second-moment noise statistics, which enable a principled correction of Rician noise-induced bias. Additionally, we introduce adaptive weighting derived from the corresponding first- or second-order moment model to normalize the variance across different signal levels, thereby mitigating the heteroscedasticity issue.

For clarity, we illustrate the proposed loss functions using the DIP framework. We assume the noisy dMRI data to be a 4D tensor $\boldsymbol{Y} \in \mathbb{R}^{m \times n \times k \times l}$,  where $m \times n \times k$ denotes the spatial dimension and $l$ represents the number of DW volumes together with the non-DW volumes. The underlying clean dMRI data is denoted as $\boldsymbol{X} \in \mathbb{R}^{m \times n \times k \times l}$, and the DIP network output (i.e., the estimation of $\boldsymbol{X}$) is represented as $\hat{\boldsymbol{X}} =f_{\hat{\boldsymbol{\theta}}}(Z)$.

\subsubsection{The first-moment noise-corrected model}
\label{sec3sub1subsubsec1}
We define the mean and variance of the observed dMRI data $\boldsymbol{Y}$ as follows:
\begin{equation}
\mu_{M1} (\boldsymbol{X_i},\sigma) = E(\boldsymbol{Y}_i |\boldsymbol{X}_i=x;\sigma)
\end{equation}
and
\begin{equation}
s_{M1}^2(\mathbf{X}_i, \sigma) = Var(\boldsymbol{Y}_i |\boldsymbol{X}_i=x;\sigma)
\end{equation}
where the closed forms under the Rician model are provided in Eq.\eqref{Eq:expectation_of_y} and \eqref{Eq:variance_of_y}. $\boldsymbol{Y}_i$ represents the signal from the $i$-th voxel in the 4D noisy dMRI dataset $\boldsymbol{Y}$, where $i=1,2,\ldots,m \times n \times k \times l$. We then fit the observation $\boldsymbol{Y}_i$ by matching it to its model-implied mean (i.e., its first moment), while standardizing the residuals with an adaptive weight proportional to the inverse of the standard deviation. This process optimizes the entire 4D dMRI data.

The loss function is given by:
\begin{equation}
\begin{split}
\mathcal{L}_{\text{DIP-M1-W1}}(\boldsymbol{\theta}) = \sum_i \left( \frac{\mu_{M1}({\hat{\boldsymbol{X}}}_i,{\sigma}) - \boldsymbol{Y}_i}{s_{M1}({\hat{\boldsymbol{X}}}_i,{\sigma})} \right)^2 \\
= \left\| \left( \mu_{\text{M1}}(\hat{\boldsymbol{X}}, \sigma) - \boldsymbol{Y} \right) \odot s_{M1}^{-1}(\hat{\boldsymbol{X}}, \sigma) \right\|_2^2
\end{split}
\end{equation}
where $\hat{\boldsymbol{X}}=f_{\hat{\boldsymbol{\theta}}}(\boldsymbol{Z})$, the symbol $\odot$ represents the tensor inner product, i.e., the element-wise product, and $\| \cdot \|_2^2$ denotes the $\ell_2$ -norm. We refer to this model as DIP-M1-W1, which adopts first-moment estimation with adaptive weighting in the DIP framework.

\subsubsection{The second-moment noise-corrected model}
\label{sec3sec1subsubsec2}
Given the straightforward nature of the second-moment formulation, we also introduce a second-moment loss function to effectively correct for both bias and variance. Specifically, we define the expectation of $\boldsymbol{Y}^2$ (second-moment) and the variance of $\boldsymbol{Y}^2$ as:
\begin{equation}
\mu_{M2} (\boldsymbol{X_i},\sigma) = E({\boldsymbol{Y}_i}^2 |\boldsymbol{X}_i=x;\sigma)
\end{equation}
and
\begin{equation}
s_{M2}^2(\mathbf{X}_i, \sigma) = Var({\boldsymbol{Y}_i}^2 |\boldsymbol{X}_i=x;\sigma)
\end{equation}
where the closed-form expressions under the Rician model are provided in Eq.\eqref{Eq:expectation_of_y^2} and \eqref{Eq:variance_of_y^2}.

The loss function is given by:
\begin{equation}
\begin{split}
\mathcal{L}_{\text{DIP-M2-W2}}(\boldsymbol{\theta}) 
= \sum_i \left( \frac{\mu_{M2}({\hat{\boldsymbol{X}}}_i,{\sigma}) - {\boldsymbol{Y}_i}^2}{s_{M2}({\hat{\boldsymbol{X}}}_i,{\sigma})} \right)^2 \\
= \left\| \left( \mu_{\text{M1}}(\hat{\boldsymbol{X}}, \sigma) - {\boldsymbol{Y}}^2 \right) \odot s_{M2}^{-1}(\hat{\boldsymbol{X}}, \sigma) \right\|_2^2
\end{split}
\end{equation}
where$\hat{\boldsymbol{X}}=f_{\hat{\boldsymbol{\theta}}}(\boldsymbol{Z})$. We refer to this model as DIP-M2-W2, which adopts second-moment estimation with adaptive weighting in the DIP framework.

\subsection{Network structure}
\label{sec3subsec2}
\begin{figure}[htbp]
\centering
\includegraphics[width=1\textwidth]{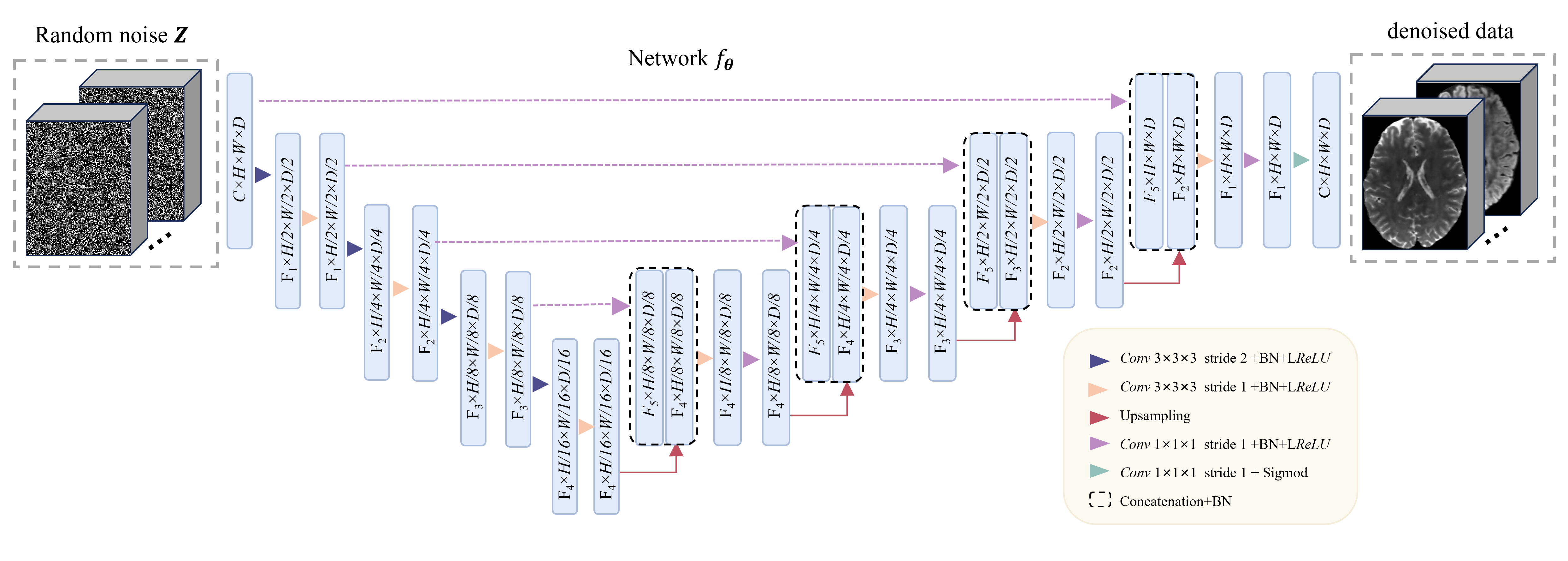}
\caption{Overview of DIP-M1-W1 and DIP-M2-W2 architectures based on 3D U-Net.}\label{Fig. 1}
\end{figure}

This study employs a 3D U-Net with a symmetric encoder-decoder architecture and skip connections. The encoder consists of multiple downsampling stages, each containing two convolutional blocks and a skip connection. The first block applies a 3×3×3 convolution with stride of 2 for downsampling, followed by batch normalization (BN) and a leaky rectified linear unit (LReLU) activation. The second block uses a 3×3×3 convolution with stride of 1, also followed by BN and LReLU. The output from the first block of each encoder stage is passed to the corresponding decoder stage through a skip connection. Before concatenation, this skip connection feature map is adjusted through a 1×1×1 convolutional layer with BN and LReLU. The decoder consists of corresponding upsampling stages. At each stage, the feature map from the previous decoder layer is upsampled and concatenated with the processed skip connection feature map. The concatenated output then passes through BN and two convolutional blocks: a 3×3×3 convolution and a 1×1×1 convolution, both with a stride of 1 and each followed by BN and LReLU. Finally, an additional 1×1×1 3D convolutional layer with a sigmoid activation outputs the denoised dMRI data (see \cref{Fig. 1}).

\subsection{Implementation details}
\label{sec3subsec3}
The two proposed algorithms were implemented using the PyTorch library. Training was conducted on a server equipped with an Intel Xeon E5-2680 CPU and a NVIDIA GeForce GTX 1080 Ti GPU. The Adam optimizer was used with an initial learning rate of 0.01, which decayed by a factor of 0.9 every 2000 iterations. The input noise map, generated from a Gaussian distribution $Z \sim \mathcal{N}(0, 1)$, matched the size of the noisy data. To achieve optimal performance and ensure a fair comparison, all DIP-based algorithms were trained for 50,000 iterations, with model parameters saved every 40 iterations. The optimal model was selected based on the training performance: for the simulated data, the output with the highest Peak Signal-to-Noise Ratio (PSNR) was chosen, while for in-vivo data, the best result was selected based on both model convergence and visual inspection. For the simulated data, the noise level ($\sigma$) is set to the ground truth. For the in-vivo data, the noise level is estimated from background.

\subsection{Simulated data}
\label{sec3subsec4}
The simulated data were generated based on the dMRI data of a single subject (“100206”) from the Human Connectome Project (HCP) WuMinn database. The data were acquired at $1.25~\text{mm}$ isotropic resolution with three distinct b-values: 18 $b = 0$ volumes, 270 DW volumes (90 volumes each at $b = 1000~\text{s/mm}^2$, $b = 2000~\text{s/mm}^2$, and $b = 3000~\text{s/mm}^2$). Acquisition was performed on a customized 3T Siemens "Connectome Skyra" scanner, with imaging parameters: TR/TE = $5520/89.5 ~\text{ms}$, matrix size = $168 \times 144 \times 111$, multi-band acceleration factor = 3, and Phase Partial Fourier = 6/8. More detailed imaging parameters are referred to \citep{Sotiropoulos2013HCPAdvances}.

The simulated data were synthesized through the following three steps \citep{Guo2022JointFramework}. First, the reference non-DW volume and diffusion tensor field were estimated by weighted linear least squares (WLLS) \citep{Veraart2013WeightedLLS}, using all $b = 0$ and $b = 1000~\text{s/mm}^2$ volumes that underwent minimal preprocessing through HCP pipelines. Second, the reference dMRI data (one non-DW volume and 30 DW volumes at $b = 1000~\text{s/mm}^2$ were reconstructed based on DTI model and then normalized into [0, 1] range. Finally, Rician noise was added to obtain the final magnitude images as follows:
\begin{equation}
\boldsymbol{Y} = \sqrt{ (\boldsymbol{X} + {\boldsymbol{N}}_1)^2 + (\boldsymbol{N}_2)^2 }
\end{equation}
where $\boldsymbol{X}$ is the noise-free dMRI data, $\boldsymbol{Y}$ is the magnitude noisy dMRI data, $\boldsymbol{N}_1$ and $\boldsymbol{N}_2$ represent independent Gaussian noise with zero mean and SD $\sigma$. The noise levels ($\sigma$) were set to 0.03, 0.05, 0.07, and 0.09.

To further evaluate the method’s ability to handle non-uniform noise distributions, reflecting a more realistic scenario in practical applications, we applied a noise pattern where the noise level ($\sigma$ map) gradually decreased from the center towards the outer regions \citep{Guo2022JointFramework}. Three spatial noise ranges were used: 0.03–0.05, 0.05–0.07, and 0.07–0.09.

\subsection{In-vivo data}
\label{sec3subsec5}
The first in-vivo data were from a published paper \citep{Moeller2021NORDIC} for denoising DW images. The data were obtained from a healthy volunteer on a 3T Siemens Magnetom Prisma scanner (Siemens Healthcare, Erlangen, Germany) with a 32-channel head coil at the Center for Magnetic Resonance Research (CMRR), University of Minnesota. Three isotropic resolutions of $1.5~\text{mm}$, $1.17~\text{mm}$ and $0.9~\text{mm}$ were acquired. Data acquisition and image reconstruction were performed with CMRR distributed C2P multiband diffusion sequence (https://www.cmrr.umn.edu/multiband/). For each resolution, a two-shell diffusion gradient scheme was employed with 7 $b = 0$ volumes, 46 directions at $b = 1500~\text{s/mm}^2$ and 46 directions at $b = 3000~\text{s/mm}^2$. Data were acquired twice, alternating the phase encoding direction (AP/PA). Each resolution used 6/8 partial Fourier with the following parameters: for $1.5 ~\text{mm}$ resolution, TR/TE = $89.2/3230 ~\text{ms}$, matrix size = $140 \times 140 \times 92$, multi-band = 4, acceleration factor = 1; for $1.17 ~\text{mm}$ resolution, TR/TE = $77.8/2780 ~\text{ms}$, matrix size = $180 \times 180 \times 120$, multi-band = 5, acceleration factor = 2; for $0.9 ~\text{mm}$ resolution, TR/TE = $95.4/5850 ~\text{ms}$, matrix size = $234 \times 234 \times 152$, multi-band = 4, acceleration factor = 2. 

The second in-vivo data was acquired from a healthy volunteer on a 3T GE SIGNA Premier scanner (GE Medical Systems, Milwaukee, America) using a 48-channel head coil at Tiantan Hospital, China \citep{Chen2025SphericalDTI}. Multiplexed sensitivity-encoding (MUSE) was used for data acquisition and reconstruction. Diffusion encoding was applied along 12 directions at $b = 800~\text{s/mm}^2$. Additionally, one $b = 0$ volume was acquired. The imaging parameters were as follows: TR/TE = $22200/66 ~\text{ms}$, matrix size = $256 \times 256 \times 150$ , in-plane resolution = $0.875~\text{mm} \times 0.875~\text{mm}$, slice thickness =$0.9 ~\text{mm}$, number of shots = 4.

The third in-vivo data was acquired from a single subject ("MGH\_1001") of the MGH-USC Human Connectome Project on a 3T Siemens Magnetom Skyra scanner (Siemens Healthineers, Germany) equipped with a 64-channel head-neck coil. Data acquisition and image reconstruction were performed using a mono-polar Stejskal-Tanner pulsed gradient spin-echo echo planar imaging (EPI) sequence with parallel imaging using Generalized Autocalibrating Partially Parallel Acquisition (GRAPPA). The data included 512 DW volumes (64 directions at $b = 1000~\text{s/mm}^2$, 64 directions at $b = 3000~\text{s/mm}^2$, 128 directions at $b = 5000~\text{s/mm}^2$, and 256 directions at $b = 10000~\text{s/mm}^2$), along with 40 non-diffusion-weighted images. Other imaging parameters were as follows: TR/TE = $8800/57 ~\text{ms}$, matrix size = $140 \times 140 \times 96$, resolution = $1.5 ~\text{mm} \times 1.5~\text{mm} \times 1.5 ~\text{mm}$, and phase partial Fourier = 6/8. More detailed imaging parameters are referred to \citep{Fan2016MGHConnectome}.

\subsection{Compared methods}
\label{sec3subsec6}
We compared the proposed methods with one widely used and three advanced denoise algorithms, including MPPCA, Patch2Self, DDM2, and Replace2Self. The introductions and relevant parameter settings are as follows.

\textbf{MPPCA} \citep{Veraart2016MPPCA}. A widely used dMRI data denoising method. Its basic idea is to project the data onto a sparse basis representation and use Random Matrix Theory to automatically identify the principal components associated with pure noise. In this work, MPPCA is from Dipy, all parameters are set to default.

\textbf{Patch2Self} \citep{Fadnavis2020Patch2Self}. An unsupervised learning dMRI data denoising method. It assumes that the noise is statistically independent across the 3D volumes of 4D dMRI data and utilize independent volumes for denoising. In this work, Pacth2Self is from Dipy, the patch radius is set to 1 and other parameters are set to default.

\textbf{DDM2} \citep{Xiang2023DDM2}. A self-supervised learning method based on the diffusion model for dMRI data. It achieves noise reduction through noise estimation, forward diffusion state matching, and reverse diffusion reconstruction. In this study, we used the code available from the website referenced in the original paper and kept all parameters at their default values.

\textbf{Replace2Self} \citep{Wu2025Replace2Self}. A novel self-supervised denoising method for dMRI data. Its principle is voxel replacement based on q-space similar block matching, combined with a complementary mask strategy, to construct image pairs for self-supervised training. In this study, we used the code provided in the original paper and kept all parameters at their default values.

As most existing methods typically adopt a two-step strategy, where Rician bias is removed after denoising, we follow previous studies and apply Rician bias correction to the denoised results of the compared methods to ensure a fair comparison\citep{Chen2025SphericalDTI}.

\subsection{Diffusion MRI metrics}
\label{sec3subsec7}
To further evaluate the denoising performance, the study conducted Diffusion Tensor Imaging (DTI) and Soma and Neurite Density Imaging (SANDI) metrics estimations based on the denoising results.

\textbf{DTI} \citep{Mori2006DTIPrinciples}. A model for water diffusion in biological tissues is based on the assumption of a Gaussian distribution per voxel and described by a $3 \times 3$ tensor. This tensor yields rotationally invariant metrics like Fractional Anisotropy (FA) and Mean Diffusivity (MD), which reflect tissue microstructure. In this work, the tensor was estimated using the WLLS from the Dipy library, with all parameters set to default.

\textbf{SANDI} \citep{Palombo2020SANDI}. A model for analyzing intricate tissue microstructure through a non-Gaussian, multi-compartment approach, which separately quantifies signal contributions from neuronal soma, neurites, and extracellular space. It provides specific metrics including neurite density $f_{in}$ (related to axon/dendrite density, high in white matter), soma fraction $f_{is}$ (indicating neuronal cell body density, high in gray matter, low in white matter), and mean soma radius $R_{s}$ (estimating neuronal soma size). In this work, we implemented the SANDI model using its official implementation, with all parameters kept at default values.

\section{Results}
\label{sec4}
\subsection{Comparison of Results for Simulated Data}
\label{sec4subsec1}
\begin{figure}[htbp]
\centering
\includegraphics[width=1\textwidth]{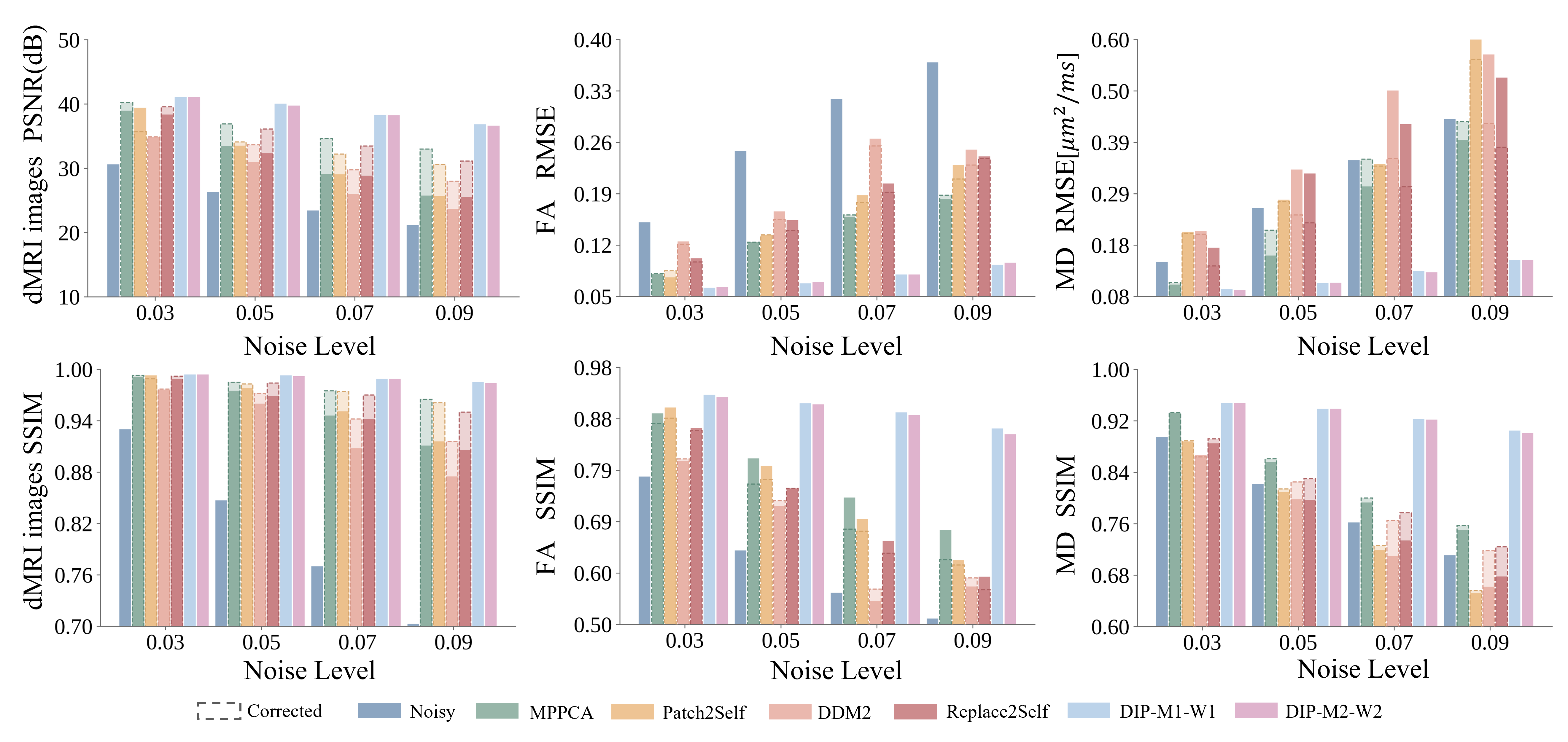}
\caption{Quantitative comparisons of different methods on the simulated dataset across various noise levels, for the denoised dMRI data and the derived DTI parameters (FA and MD). Dashed lines represent denoised results after applying explicit first-moment Rician bias correction.}\label{Fig. 2}
\end{figure}

\cref{Fig. 2} presents the PSNR and Structural Similarity Index (SSIM) of the denoised dMRI data produced by our methods and compared methods (with and without first-moment Rician bias correction) at various noise levels, together with the quantitative evaluation of the corresponding DTI parametric maps (FA and MD) using the Root-Mean-Square Error (RMSE) and SSIM. For the compared methods, Rician bias correction significantly improved the quantitative metrics (PSNR and SSIM) of the DW images across all noise level. However, this enhancement appears to compromise the precision of the derived diffusion metrics. Specifically, the RMSE increased significantly for MPPCA-corrected MD maps, and the SSIM decreased notably for the FA maps of all methods except DDM2. In contrast, our methods (both DIP-M1-W1 and DIP-M2-W2) consistently outperforms the others across all noise levels, providing higher PSNR and SSIM for the denoised images and lower RMSE along with higher SSIM for the FA and MD maps.

\begin{figure}[htbp]
\centering
\includegraphics[width=1\textwidth]{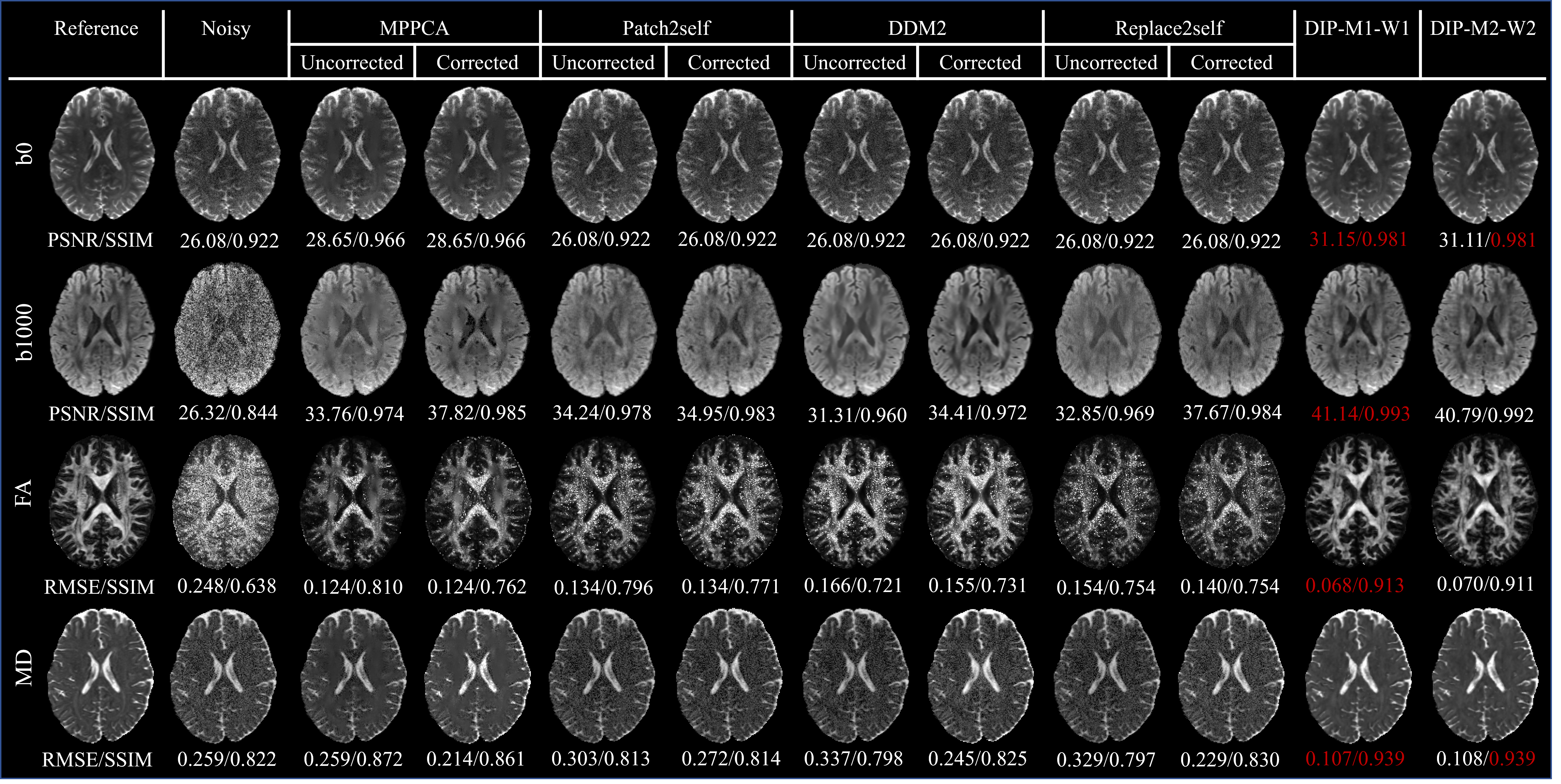}
\caption{Visual comparison of non-DW (b0) and DW (b1000) images, as well as FA and MD maps, obtained by our methods and by competing methods (both with and without Rician bias correction) ,on the simulated data with a noise level of 0.05. The values highlighted in red indicate the best performance.}\label{Fig. 3}
\end{figure}

As shown in \cref{Fig. 3}, while methods such as MPPCA, Patch2Self, DDM2, and Replace2Self reduce noise fluctuation, they do not inherently correct the underlying noise bias, limiting their performance—particularly in low-SNR regions like the cerebrospinal fluid. When corrected by Rician noise correction, these denoised results exhibit a reduction in global brightness, bringing the DW images closer to the ground truth and thus improving visual fidelity. In contrast, both DIP-M1-W1 and DIP-M2-W2 substantially reduce noise while correcting for bias and preserving fine structural details, producing results that are visually closest to the ground truth among all compared methods for the dMRI images and DTI parametric maps. 

In summary, while explicit noise correction can improve the compared methods’ performance to some extent, both the quantitative metrics and the visual assessments indicate that they still fall short of the results achieved by our proposed approach. Based on the above results and to ensure a fair comparison, only the noise-corrected results of the comparison methods are reported in subsequent experiments.

\begin{figure}[htbp]
\centering
\includegraphics[width=1\textwidth]{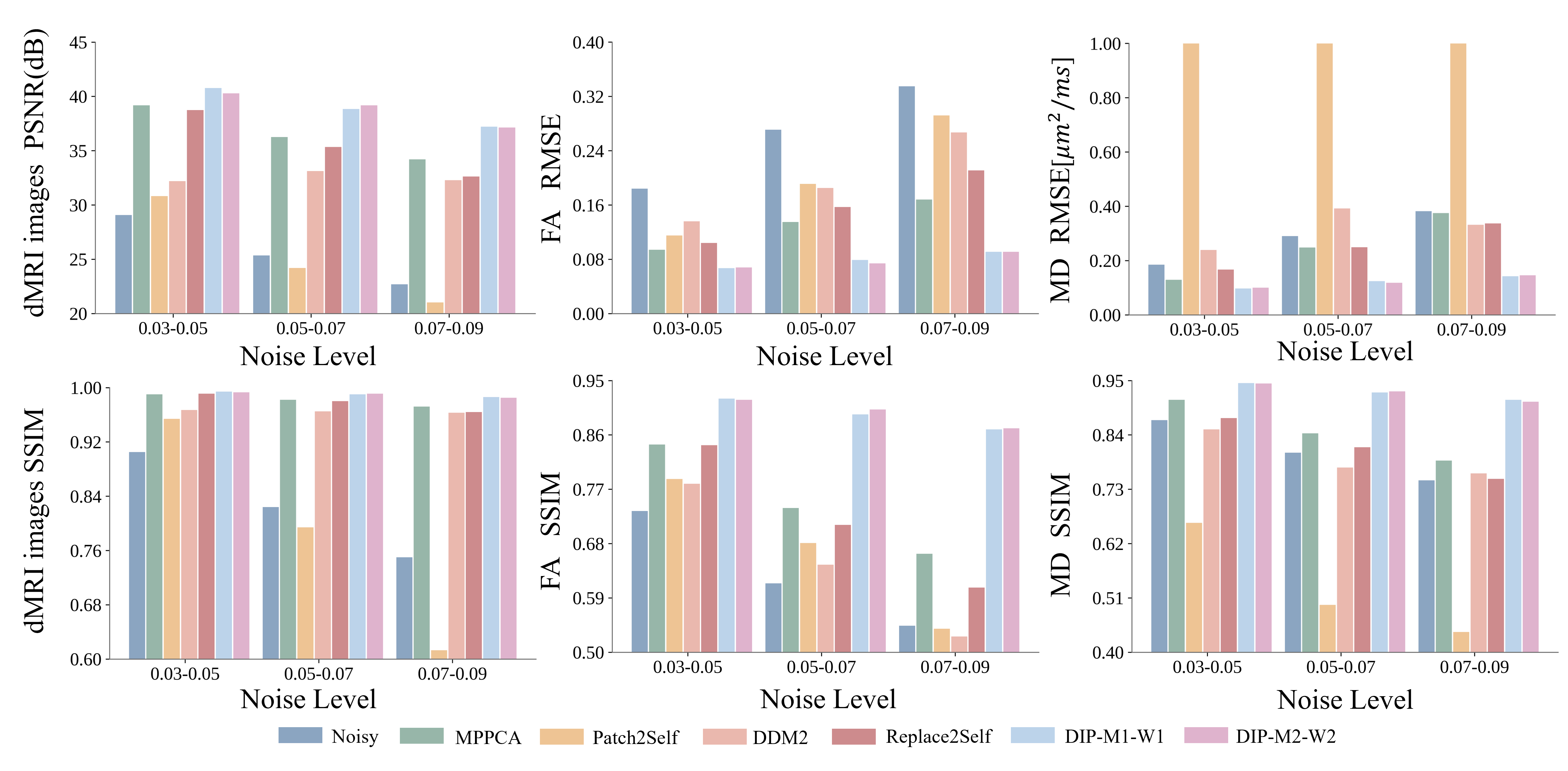}
\caption{Quantitative evaluation of different methods on simulated dMRI data with spatially varying noise: PSNR and SSIM of dMRI images, and RMSE and SSIM of FA and MD maps.}\label{Fig. 4}
\end{figure}

To better reflect the realistic scenario of spatially varying noise in practical applications, we further evaluate the capability of our proposed method for handling non-uniform noise by incorporating a voxel-wise noise map in the denoising process. As shown in \cref{Fig. 4}, the PSNR and SSIM of the denoised results obtained by different methods under different spatially varying noise are presented, along with the RMSE and SSIM results for FA and MD. Experimental results demonstrate that under various non-uniform noise conditions, our proposed method (DIP-M1-W1 and DIP-M2-W2) consistently outperforms all compared methods. Specifically, it achieves higher PSNR and SSIM for the denoised images, and delivers more accurate reconstruction of diffusion metrics (FA and MD), as reflected by lower RMSE and higher SSIM. 

\begin{figure}[htbp]
\centering
\includegraphics[width=1\textwidth]{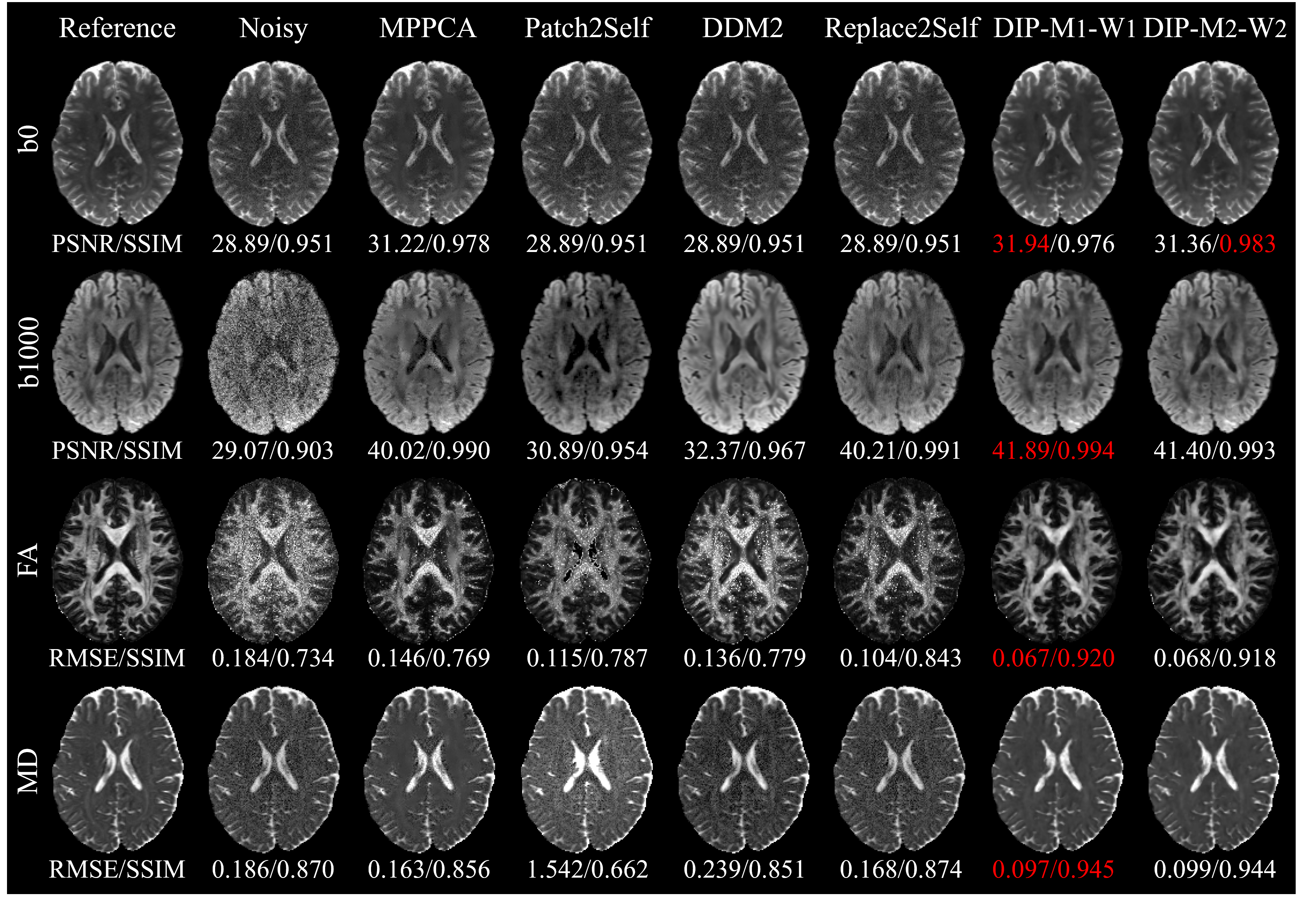}
\caption{Visual comparison of non-DW (b0) and DW (b1000) images, as well as FA and MD maps, obtained by different methods using the simulated data with spatially varying noise (noise level 0.03-0.05). The values highlighted in red indicate the best performance.}\label{Fig. 5}
\end{figure}

As shown in \cref{Fig. 5}, while methods such as MPPCA, DDM2, and Replace2Self achieve some reduction in noise fluctuations and deviations after correction, their overall performance is fundamentally constrained by their sequential processing pipelines, where errors accumulate across steps. In contrast, Patch2Self tends to over-correct, resulting in an underestimation of the dMRI signal and, consequently, a substantial deviation of the MD values from the ground truth, as evidenced by its high RMSE. Unlike these separate-step methods, our proposed DIP-M1-W1 and DIP-M2-W2 frameworks integrate bias correction seamlessly into the denoising process. This unified approach not only achieves effective noise suppression but also preserves fine structural details, demonstrating clear visual superiority. In summary, the proposed method is capable of highly effective denoising performance against spatially non-uniform noise by using the provided voxel-wise noise map ($\sigma(x)$).

\subsection{Comparison of Results for In-vivo data}
\label{sec4subsec2}
\begin{figure}[htbp]
\centering
\includegraphics[width=1\textwidth]{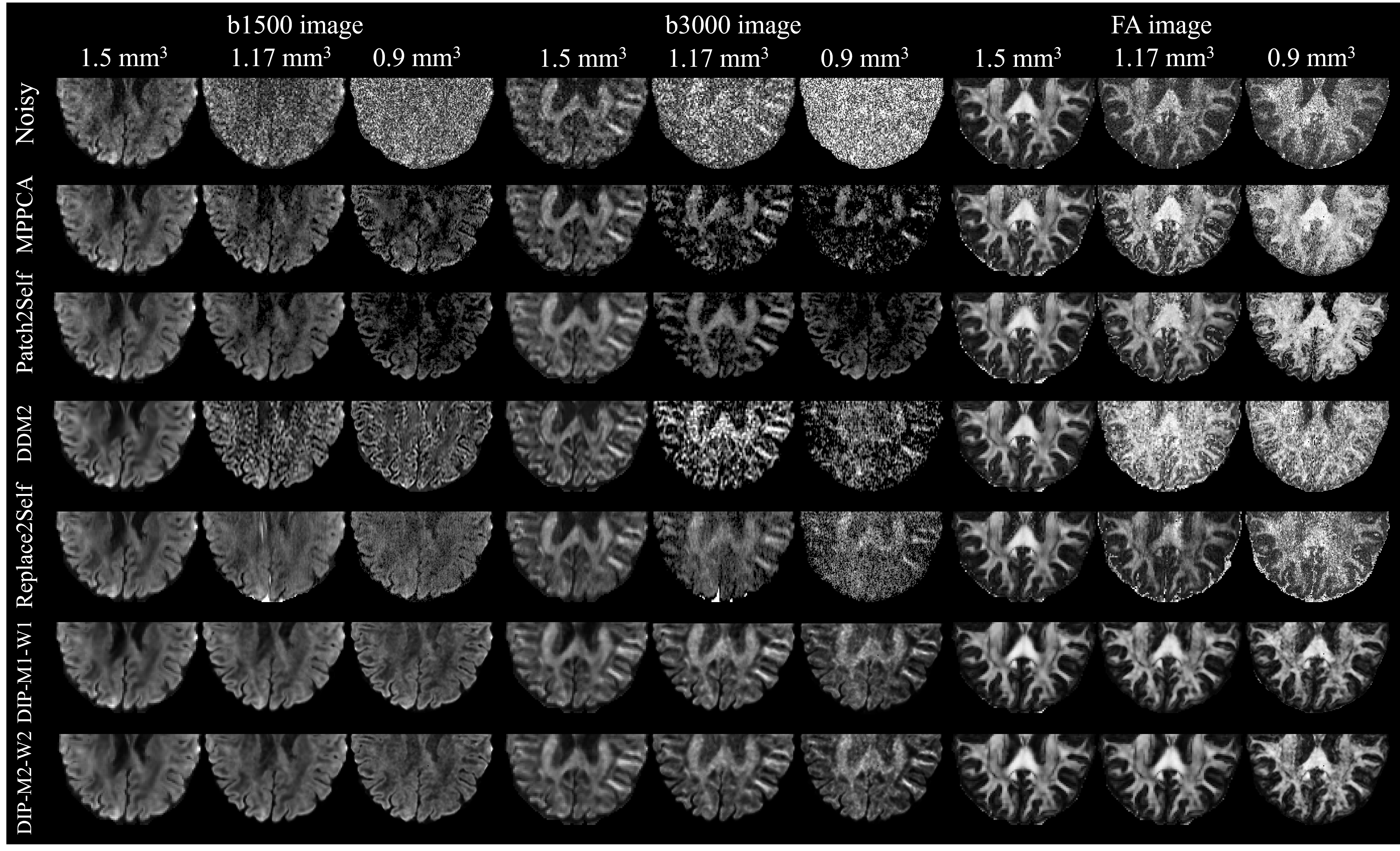}
\caption{Visual comparison of DW images ($b = 1500$ and $ 3000~\text{s/mm}^2$) and FA maps from different methods, using the first in-vivo data with three different resolutions.}\label{Fig. 6}
\end{figure}

In \cref{Fig. 6}, we compare different denoising algorithms using the first in-vivo data at different resolutions. For the $1.5 ~\text{mm}$ resolution data, with high SNR, all denoising algorithms can effectively remove noise while preserving structural details. In the $1.17 ~\text{mm}$ high-resolution data, noise severely disrupts the primary structure of the original image, particularly pronounced in the DW image with $b = 3000~\text{s/mm}^2$. Following Rician bias correction, methods such as MPPCA, Patch2Self, and DDM2 effectively remove noise and Rician bias, but simultaneously cause an undesirable attenuation of the dMRI signal. This signal attenuation leads to a loss of structural details in the DW images, compromising overall denoising performance and propagating errors into FA map estimation. Moreover, DDM2 introduces spurious structural features. For the $0.9 ~\text{mm}$ data, which exhibits further SNR reduction, the raw images show almost unrecognizable internal brain structures. Consistent with observations from the $1.17 ~\text{mm}$ data, all compared denoising methods showed limited efficacy for $0.9 ~\text{mm}$ data. In contrast, the proposed DIP-M1-W1 and DIP-M2-W2 methods significantly reduce noise, producing high-quality images with clearer details, less residual noise, and greatly enhanced visibility of anatomical structures.

\begin{figure}[htbp]
\centering
\includegraphics[width=1\textwidth]{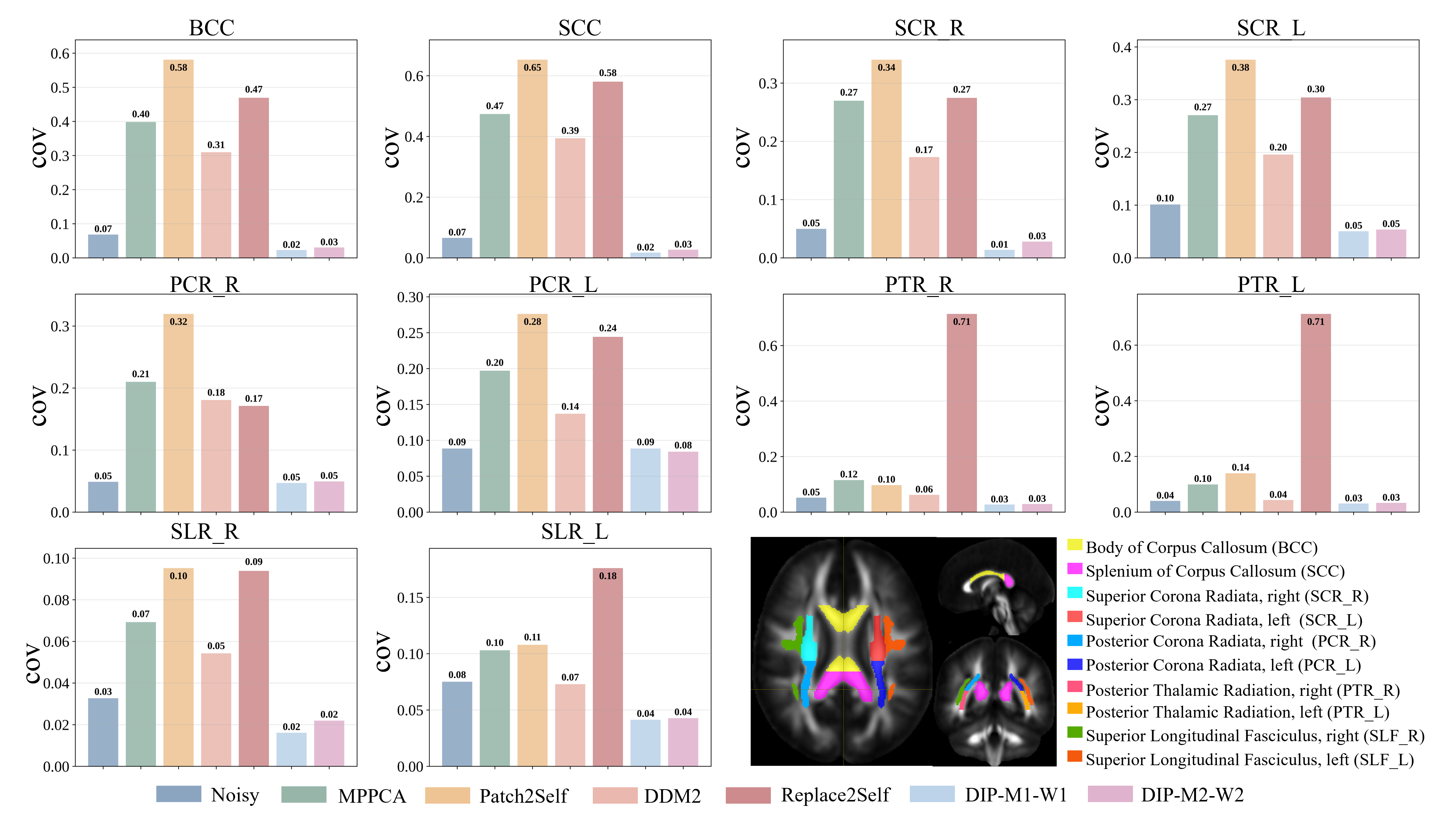}
\caption{Comparison of the coefficient of variation (COV) for FA values across different brain regions from different methods, using the first in-vivo data with three different resolutions.}\label{Fig. 7}
\end{figure}

Furthermore, as a biomarker indicative of the inherent properties of white matter (WM) microstructure, FA should remain consistent across different image resolutions for a given brain region within the same subject. Consequently, we provide the Coefficient of Variation (COV) of FA values across three resolutions to illustrate the stability of different denoising methods (see \cref{Fig. 7}). Specifically, the denoised data from three resolutions were first registered to a standard white matter parcellation template. For each brain region, the mean FA was calculated separately for each of the three resolutions, and the COV was computed across the three mean FA values. As shown in \cref{Fig. 7}, the COV values for MPPCA, Patch2Self, DDM2, and Replace2Self all maintain relatively high levels across all brain regions, indicating that these methods fail to maintain stable FA metrics across different resolutions. The proposed method achieved the lowest COV in all regions. In summary, the proposed DIP-M1-W1 and DIP-M2-W2 methods consistently deliver exceptional denoising performance across data of varying resolutions, demonstrating outstanding and robust performance.

\begin{figure}[htbp]
\centering
\includegraphics[width=1\textwidth]{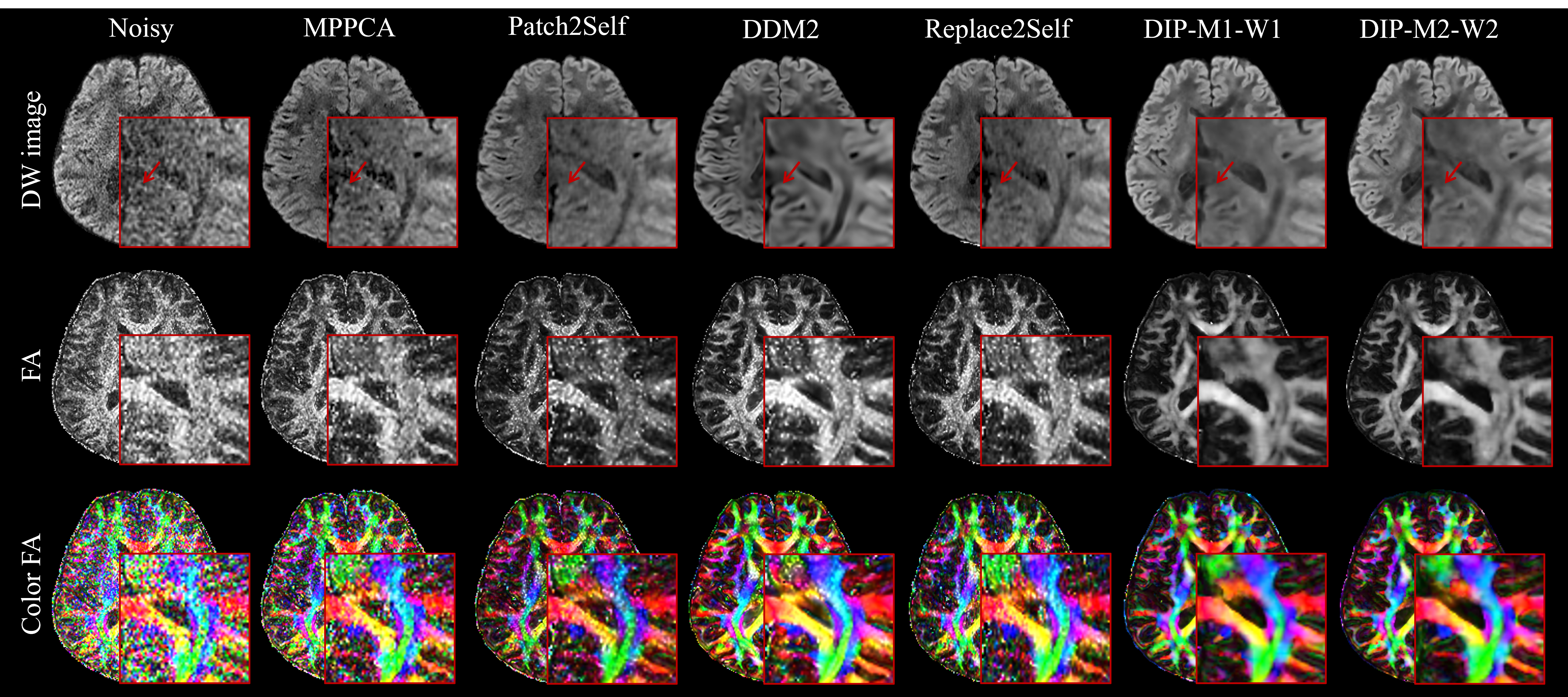}
\caption{A visual comparison of DW image, FA, color-FA, and their corresponding enlarged maps for different methods, using the second in-vivo data.}\label{Fig. 8}
\end{figure}

In \cref{Fig. 8}, we compare different denoising algorithms using the second in-vivo data with high-spatial resolution. The results generated by MPPCA, Patch2Self, and Replace2self are suboptimal, likely due to the limited number of diffusion directions (12 directions) in this dataset, which restricts the availability of directional correlation information necessary for effective denoising. Although DDM2 suppresses noise effectively, producing smooth results, its multi-stage learning may generate false structures (see red arrows in \cref{Fig. 8}). In contrast, the DIP-M1-W1 and DIP-M2-W2 not only effectively suppress noise but also retain a significant amount of detail information. Note that the b0 image is not denoised for Patch2Self, DDM2, and Replace2Self because their algorithmic frameworks can neither process a single b0 volume nor jointly process b0 and DW volumes. Consequently, the FA maps obtained from these methods still contain a substantial amount of noise. In summary, the proposed method DIP-M1-W1 and DIP-M2-W2 demonstrates robustness and applicability even under limited data conditions.

\begin{figure}[htbp]
\centering
\includegraphics[width=1\textwidth]{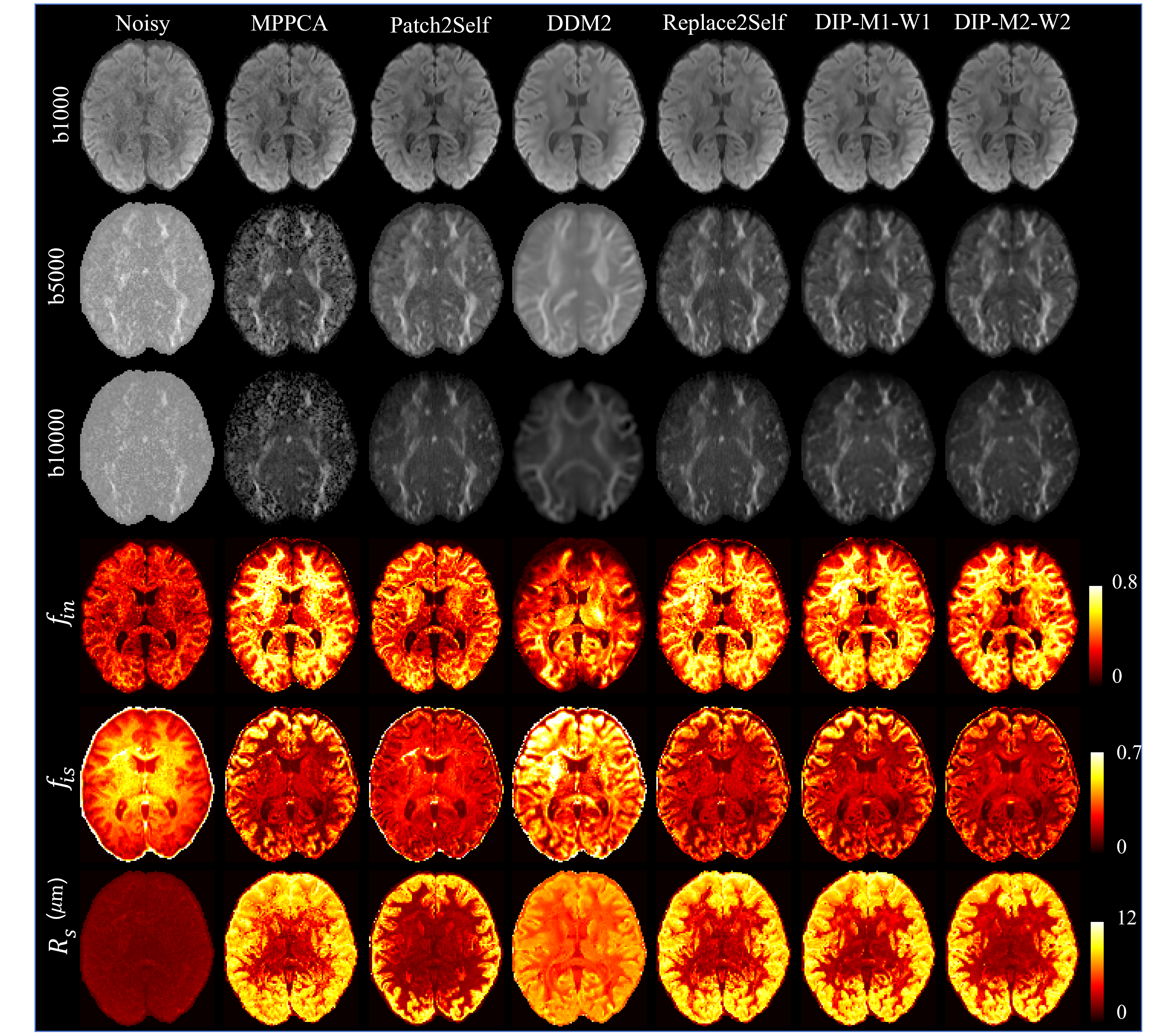}
\caption{Comparison of DW images and SANDI parameter maps ($f_{in}$,$f_{is}$, and $R_{s}$) generated by different denoising methods from the third in-vivo data with ultra-high b-values.}\label{Fig. 9}
\end{figure}

\cref{Fig. 9} illustrates the visualization of DW images and SANDI parameter maps from different denoising methods, using the third in-vivo data with ultra-high b-values. It is evident from the DW images that a pronounced noise floor is present in the high b-value data. Notably, MPPCA exhibits substantial signal attenuation due to over‑correction. In comparison, DDM2 produces overly smooth results accompanied by spurious structural artifacts, while the noise floor remains pronounced with Patch2Self. In contrast, the noise‑corrected results from Replace2Self effectively suppress noise while preserving substantial structural detail in the ultra‑high b‑value DW images. Similarly, the proposed DIP-M1-W1 and DIP-M2-W2 not only suppress noise effectively but also eliminate bias, thereby ensuring the preservation of significant details. 

To better illustrate the pronounced impact of noise on downstream diffusion metrics, we performed SANDI model fitting on the denoised data. As shown in \cref{Fig. 9}, the poor denoising performance of DDM2 yields SANDI metric maps that contradict established physiology. In contrast, although Patch2Self retains structural details in the DW images, its derived $f_{in}$ and $f_{is}$ maps fail to highlight their respective target tissues—white matter and gray matter—with the expected contrast, likely due to residual noise floor obscuring the underlying signal differentiation. Notably, both Replace2Self and our proposed methods (DIP-M1-W1 and DIP-M2-W2) generate qualitatively similar SANDI parameter maps, where the $f_{in}$ and $f_{is}$   maps clearly delineate white-gray matter boundary, and the estimated $R_{s}$ values fall within the biologically plausible range of 5–15 $\mu$m. The SANDI metrics maps from the proposed methods exhibit a higher level of anatomical rationality and physiological consistency, both visually and numerically. In summary, the proposed DIP-M1-W1 and DIP-M2-W2, by introducing noise bias and variance heterogeneity correction mechanisms, demonstrate superior performance in handling high-bias noise and extremely low SNR data. 

\subsection{Ablation results}
\label{sec4subsec3}
\subsubsection{Effect of Noise Bias and variance Correction}
\label{sec4sub3subsubsec1}
\begin{figure}[htbp]
\centering
\includegraphics[width=1\textwidth]{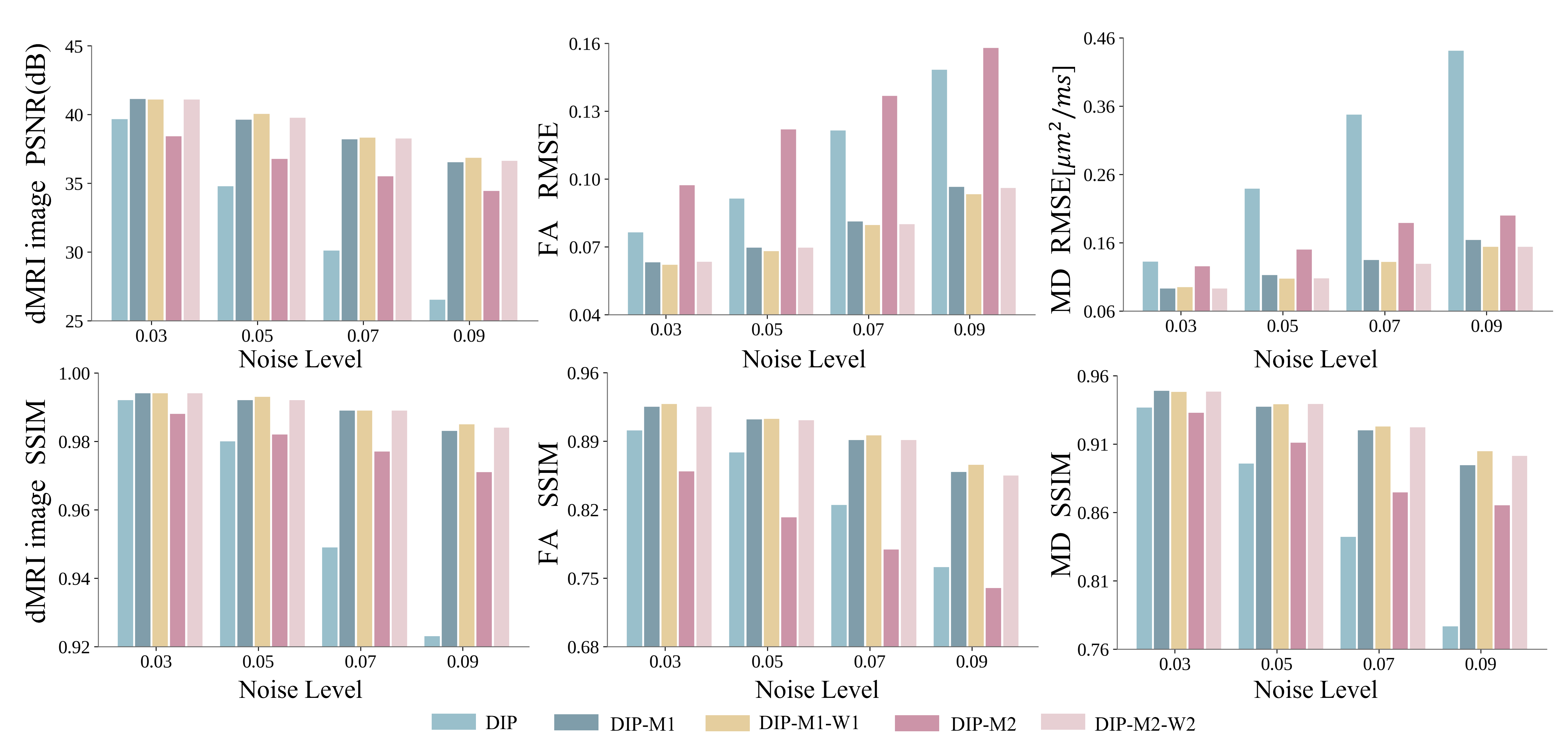}
\caption{Quantitative comparisons of different DIP-based methods on the simulated dataset with different noise levels, for the denoised dMRI data and the derived DTI parameters (FA and MD).}\label{Fig. 10}
\end{figure}

\cref{Fig. 10} illustrates the PSNR and SSIM values of denoised dMRI data from different DIP-based methods at various noise levels, together with the RMSE and SSIM of the corresponding FA and MD maps. DIP-M1-W1 slightly outperforms DIP-M1, with both methods significantly surpassing original DIP in dMRI images (PSNR, SSIM) and FA/MD maps (RMSE, SSIM). The similar performance between DIP-M1 and DIP-M1-W1 is reasonable, as the variance heterogeneity in the first-moment model is small, making the correction effect from the introduced weighting term minimal. Regarding DIP-M2, it performs better than original DIP for images and MD maps when the noise level is $\geq$ 0.05. However, its performance in the FA maps is significantly worse than DIP. By introducing weighted terms to correct for variance heterogeneity, DIP-M2-W2 significantly outperforms DIP-M2 and DIP across all noise levels. In summary, the proposed DIP-M1-W1 and DIP-M2-W2 produced higher PSNR and SSIM for the denoised images, as well as lower RMSE for the FA and MD maps, demonstrating the effectiveness of correcting biases and heteroscedasticity.

\subsubsection{Effect of Noise Estimation Accuracy}
\label{sec4sub3subsubsec2}
\begin{figure}[htbp]
\centering
\includegraphics[width=1\textwidth]{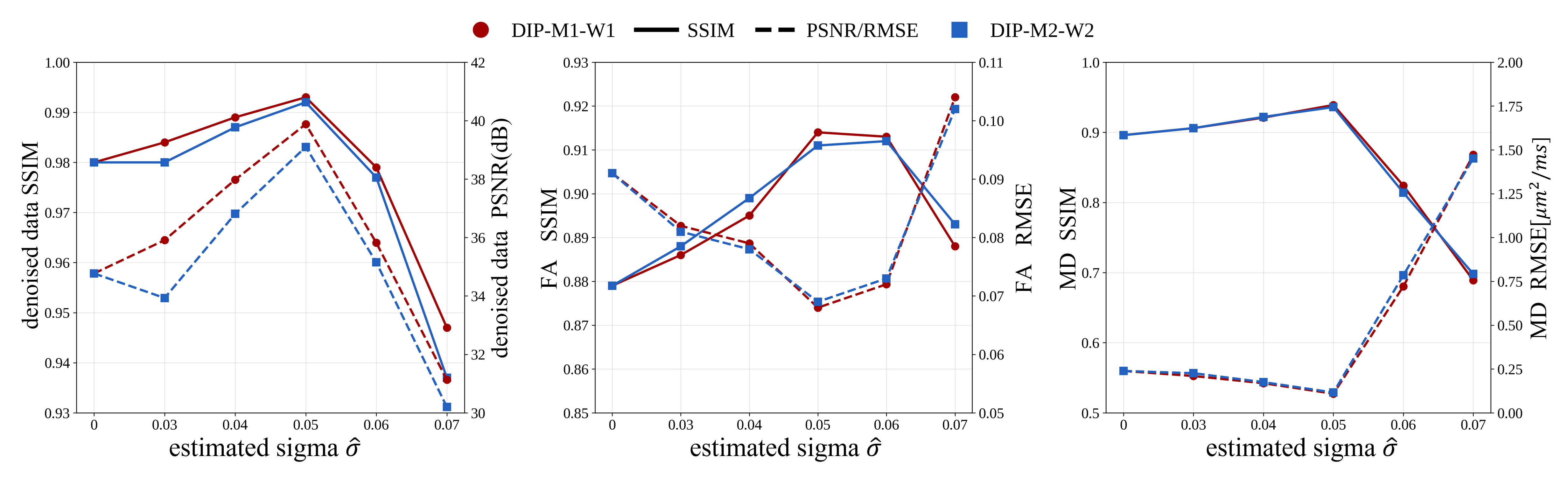}
\caption{Quantitative comparisons of DIP-M1-W1 and DIP-M2-W2 on the denoised dMRI data and derived DTI parameters (FA and MD), using different noise estimation ($\hat{\sigma}$) on the simulated data with a true noise level of 0.05 ($\hat{\sigma}=0$: the original DIP).}\label{Fig. 11}
\end{figure}

\cref{Fig. 11} displays the PSNR and SSIM of the denoised results from DIP-M1-W1 and DIP-M2-W2, as well as the RMSE and SSIM of the FA and MD, using different noise estimations ($\hat{\sigma}$= 0.03, 0.04, 0.05, 0.06, 0.07) on the simulated data with a true noise level $\sigma$ of 0.05. As illustrated in \cref{Fig. 11}, underestimating the $\sigma$ ($\sigma<\hat{\sigma} $) leads to incomplete noise correction, with the denoising performance improving as the estimated noise level approaches the true noise level. Even when the noise level is underestimated, the performance still surpasses that of the original DIP ($\hat{\sigma}=0$). However, overestimating the $\sigma$ ($\sigma>\hat{\sigma} $) causes over-correction, which underestimates the dMRI signal and yields poor denoising outcomes. The greater the deviation of the noise estimation from the actual noise level, the poorer the denoising performance. In summary, the accuracy of noise estimation has a crucial impact on the denoising capabilities of DIP-M1-W1 and DIP-M2-W2.

\begin{figure}[htbp]
\centering
\includegraphics[width=1\textwidth]{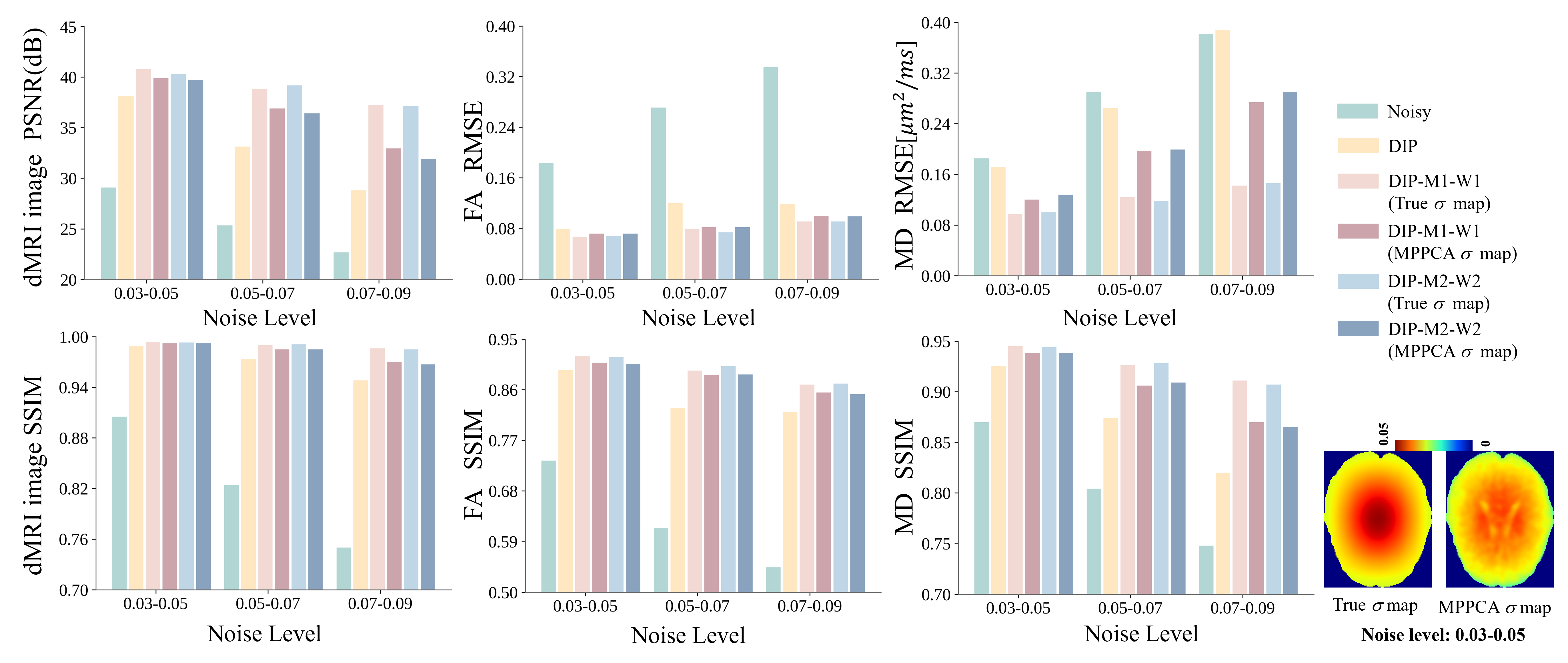}
\caption{Quantitative comparisons of DIP-M1-W1 and DIP-M2-W2 on the denoised dMRI data and derived DTI parameters (FA and MD), using both MPPCA-estimated and true noise maps on a simulated dataset with spatially varying noise. The lower right corner displays the MPPCA-estimated and true noise maps for the spatially varying noise (0.03-0.05).}\label{Fig. 12}
\end{figure}
\cref{Fig. 12} illustrates the PSNR and SSIM of denoised results from DIP-M1-W1 and DIP-M2-W2 using the true noise map and the MPPCA-estimated noise map under different spatially varying noise, along with the RMSE and SSIM results for FA and MD. At relatively low noise levels (0.03-0.05), denoising with the MPPCA-estimated noise map yields results comparable to those obtained with the true noise map (see \cref{Fig. 12} for the MPPCA-estimated and true noise map). However, as the level of spatial noise increases, the noise bias estimated by MPPCA also increases. Consequently, using the MPPCA-estimated noise map leads to a growing discrepancy from the ideal denoising result, an effect most pronounced in the MD metric. Despite inaccurate noise estimation, particularly in regions with high noise levels, the proposed method using the MPPCA-estimated noise map still significantly outperforms the uncorrected baseline (DIP). In summary, by leveraging a noise map ($\sigma(x)$) estimated from existing methods, the proposed methods can also achieve highly effective denoising performance against spatially non-uniform noise.
\begin{figure}[htbp]
\centering
\includegraphics[width=1\textwidth]{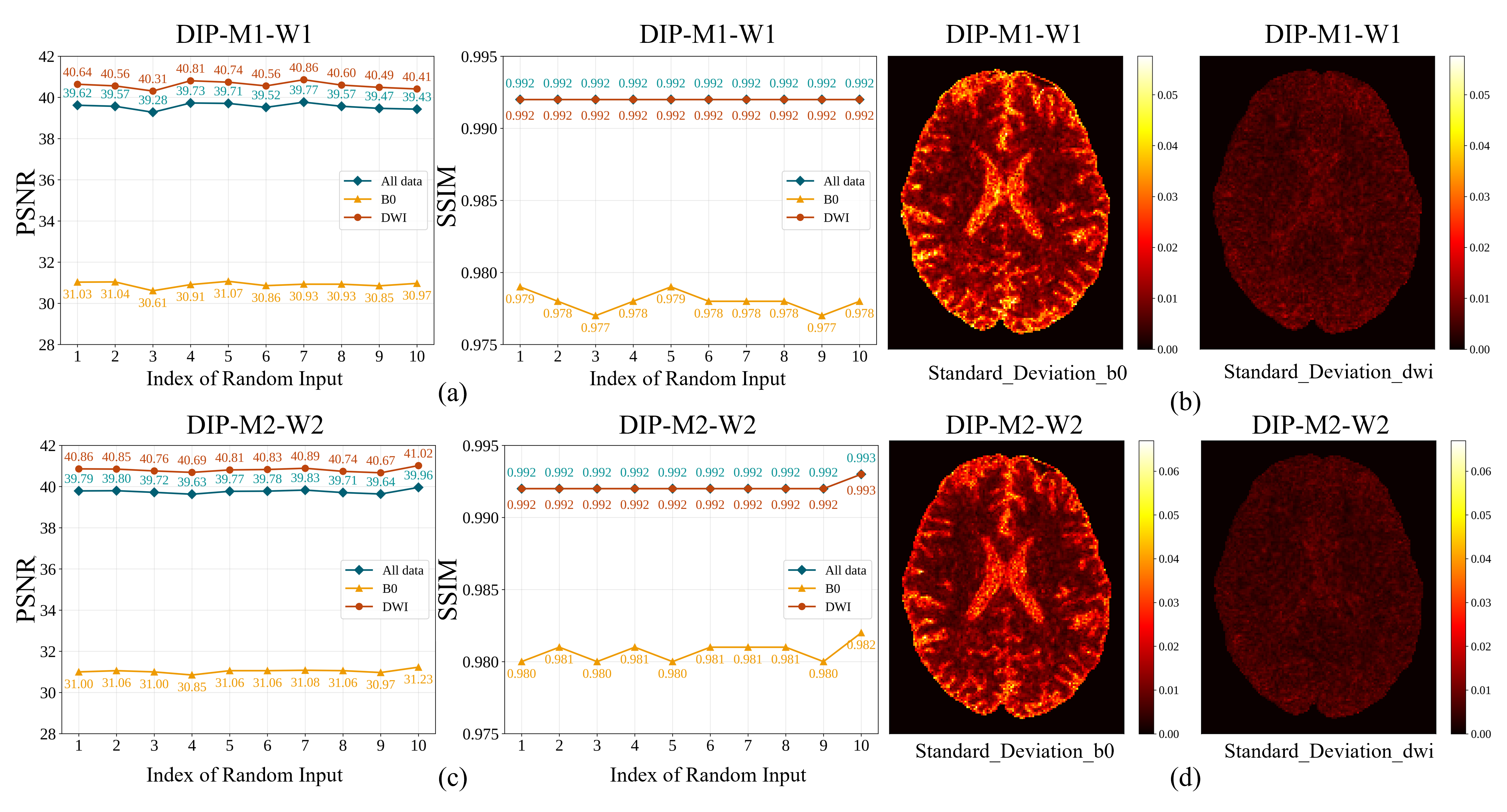}
\caption{Robustness evaluation under random initializations. (a, c) PSNR and SSIM of the denoised images across ten random initializations for DIP-M1-W1 and DIP-M2-W2, respectively. (b, d) Corresponding standard deviation maps derived from the ten denoised b0 images and DW images for DIP-M1-W1 and DIP-M2-W2.}\label{Fig. 13}
\end{figure}

\subsubsection{Robustness Test}
\label{sec4sub3subsubsec3}
\cref{Fig. 13} illustrates the robustness evaluation of DIP-M1-W1 and DIP-M2-W2, showing their denoising performance across ten random initializations. Specifically, \cref{Fig. 13} (a) and (c) show the PSNR and SSIM variation curves of the denoising results (all data, b0, and DW images) from ten random initializations for the DIP-M1-W1 and DIP-M2-W2, while \cref{Fig. 13} (b) and (d) display standard deviation maps of the ten denoise results, include the denoised b0 and DW images. The results indicate that the PSNR and SSIM curves for both methods exhibit minimal fluctuations under different random initializations, and the standard deviations of the b0 images and DW images remain at low levels. In summary, the proposed method demonstrates high stability and robustness.

\section{Discussion}
\label{sec5}
In this work, we introduce two novel loss functions, DIP-M1-W1 and DIP-M2-W2, designed for unsupervised denoising of DW images. These losses explicitly address Rician noise within the DIP framework. Through simulated and in-vivo experiments, we show that DIP-M1-W1 and DIP-M2-W2 provide accurate, unbiased results for both DW images denoising and subsequent diffusion-metric estimation, outperforming state-of-the-art methods. Central to our approach is leveraging first- or second-order moment models for bias correction and employing adaptive weighting to handle variance heterogeneity.

At high SNR, the Rician distribution approaches Gaussian and the bias becomes negligible. In the low-SNR regime typical of dMRI, however, the Rician noise floor elevates diffusion-weighted intensities \citep{Huynh2024OptimalShrinkage,Manzano2024DenoisingConsiderations}. Denoising objectives that assume i.i.d. Gaussian noise can reduce noise fluctuations but cannot remove this noise-induced bias, yielding systematic overestimation of DW signal. This upward bias propagates to downstream diffusion metrics, potentially confounding clinical interpretation. For example, in DTI it yields underestimated MD and attenuated FA, particularly in highly anisotropic tissue, because the noise floor lifts low-SNR DW measurements, making high-b signals appear less attenuated and compressing direction-dependent contrast. 

By incorporating expectation-based modeling directly into the denoising loss, the proposed approach enables theoretical correction of the Rician bias during optimization, rather than as a post-processing step. Under the first-moment loss, noise-variance heterogeneity is mild: the standard deviation ranges from $\approx 0.43\sigma$ in the zero-signal limit to approaching $ \sigma $ as SNR becomes large, i.e., a span of roughly $\sim 1.5\times$. Consistent with this, our experiments show that explicitly correcting variance heterogeneity (DIP-M1-W1) brings little additional benefit beyond only bias correction (DIP-M1). This suggests that, in practice, applying only bias correction can be effective under the first-moment loss (see \cref{Fig. 10}). In contrast, with the second-moment loss, the heterogeneity is pronounced: the standard deviation is $2\sigma\sqrt{s^2 + \sigma^2}$ (where $s$ is the ground truth), thus the heterogeneity scale is ${\sqrt{s_{\max}^2 + \sigma^2}}/{\sqrt{s_{\min}^2 + \sigma^2}} \approx s_{\max}/\sigma = \text{SNR}$  (e.g., $\sim 60$ when SNR$ \approx 20$). In this regime, jointly correcting both bias and variance is essential; otherwise, the optimization overweights high-variance voxels and severely degrades denoising quality, ultimately distorting the derived diffusion metrics.

Accurate noise-level estimation ($\sigma$) is critical for our moment-based losses, which use $\sigma$ to calibrate the bias-correction term. Mis-specifying $\sigma$—whether over- or under-estimating it — degrades denoising and quantitative metrics, and the impact typically grows with error magnitude (see \cref{Fig. 11}). Overestimation ($\sigma>\hat{\sigma} $) drives overly aggressive correction, attenuating intensities and yielding a downward bias (systematic underestimation of the true signal). By contrast, underestimation ($\sigma<\hat{\sigma}$) produces incomplete correction and is generally less detrimental. For equal-magnitude errors, for example, when the true noise level $\sigma$ is 0.05, underestimation ($\hat{\sigma}= 0.04$) causes smaller drops in DW-image PSNR and smaller increases in FA/MD RMSE than overestimation ($\hat{\sigma}= 0.06$). Thus, in practice, slightly conservative (lower) $\hat{\sigma}$ values are safer. 

For spatially invariant noise, $\sigma$ can be estimated from background (air) regions. For spatially varying noise — e.g., due to parallel imaging and receive-field modulation — our formulation remains valid when supplied with a voxel-wise noise map $\sigma(x)$ (see \cref{Fig. 12}). The spatially varying noise map $\sigma(x)$ can be estimated accurately using the g-factor when available \citep{Moeller2021NORDIC,Moeller2021PostHCP}. If a g-factor map is not provided, PCA-based estimators \citep{Veraart2016MPPCA,Manjon2013OvercompletePCA} offer an effective alternative for estimating $\sigma(x)$. Their estimates are typically slightly conservative (underestimated), which is preferable and safer for our loss than overestimation. In addition, jointly estimating $\sigma(x)$ with denoising via an embedded diffusion-MRI model (e.g., DTI) \citep{Landman2009SpatiallyVariableNoise} has the potential to improve the noise map and denoising, but it requires specifying a particular model.

DIP-style optimization is computationally expensive. For a dMRI dataset of size $180 \times 180 \times 90$ and 31 directions, denoising took  $\sim 10$ hours per dataset for all tested DIP-based algorithms under a fixed budget of 50,000 iterations. We used a fixed budget because DIP lacks a universally accepted stopping criterion, and this setting yielded stable convergence across noise levels. Despite the cost, we adopt DIP for two reasons. First, its objective is the classical least-squares data-fidelity term, which keeps the model simple, easy to implement, and straightforward to interpret, enabling a fair evaluation of the proposed bias and variance corrections. Second, DIP avoids external training data and large models, offering a clean testbed that isolates the effects of bias correction and variance modeling without attributing gains to pretraining, data curation, or architectural choices.

The first-moment-based loss (DIP-M1-W1) corrects bias by matching the noisy magnitude to its expectation modeled by true underlying signal. This requires evaluating modified Bessel functions, which makes the method more complex to implement, but optimization is typically stable. The second-moment-based loss (DIP-M2-W2) instead targets squared-signal bias by matching the second moment. It has a simple closed form and is easy to implement and interpret. However, because it operates in the squared domain, using a standard L2 residual yields an objective that is quartic in the signal, which is generally less stable than DIP-M1-W1. Understanding these trade-offs helps users choose the model that best aligns with their denoising goals and computational constraints.

The effectiveness of our bias- and variance-correction loss is verified within DIP. The noise characteristics of multi-coil magnitude MRI vary depending on the reconstruction method. For example, SENSE-type reconstructions typically produce magnitude data that follows a Rician distribution, while magnitude-combined reconstructions, such as GRAPPA, are better described by a noncentral chi distribution \citep{Varadarajan2015MajorizeMinimize}. Extending our moment-based losses to account for this broader noise model is conceptually straightforward and will be explored in future work.   Beyond DIP, the proposed bias- and variance-correction losses are portable. They can be integrated into other unsupervised or self-supervised denoising frameworks by tailoring the loss to each method’s native objective. Moreover, the same formulation is pertinent to other low-SNR MRI regimes, including low-field MRI, high-spatial-resolution acquisitions and highly accelerated imaging, as well as in functional MRI and quantitative parameter mapping (e.g., T1, T2, and T2* mapping). A systematic evaluation of these extensions is left to future work.

\section{Conclusion}
\label{sec16}
In conclusion, we proposed two novel noise-correction loss functions, embedded in the DIP framework, for accurate and unbiased magnitude dMRI denoising. Experiments on simulated and in-vivo diffusion data demonstrate effective bias and variance correction, yielding superior denoising and strong performance in downstream diffusion-parameter estimation. As such, we believe that the two proposed denoising algorithms are beneficial for achieving high-quality dMRI, which is desirable for diverse neuroscientific and clinical applications.

\section*{Acknowledgment}
This work was supported by National Natural Science Foundation of China under Grants U21A6005, 82372079, Natural Science Foundation of Guangdong Province (2024A1515010014), and Guangdong-Hong Kong Joint Laboratory for Psychiatric Disorders (2023B1212120004).

\bibliographystyle{elsarticle-harv} 
\bibliography{DIP_reference}

\end{document}